\documentclass[10pt,a4paper,hidelinks,twocolumn]{article}

\usepackage[T1]{fontenc}
\usepackage[english]{babel}
\usepackage[utf8]{inputenc}

\usepackage{graphicx}
\usepackage[table,usenames,dvipsnames]{xcolor}
\usepackage{booktabs}
\usepackage{tabularx}

\usepackage{fancyhdr}
\usepackage{pdfpages}

\fancypagestyle{CISPA}{

\fancyhead[C]{\includegraphics[width=4cm]{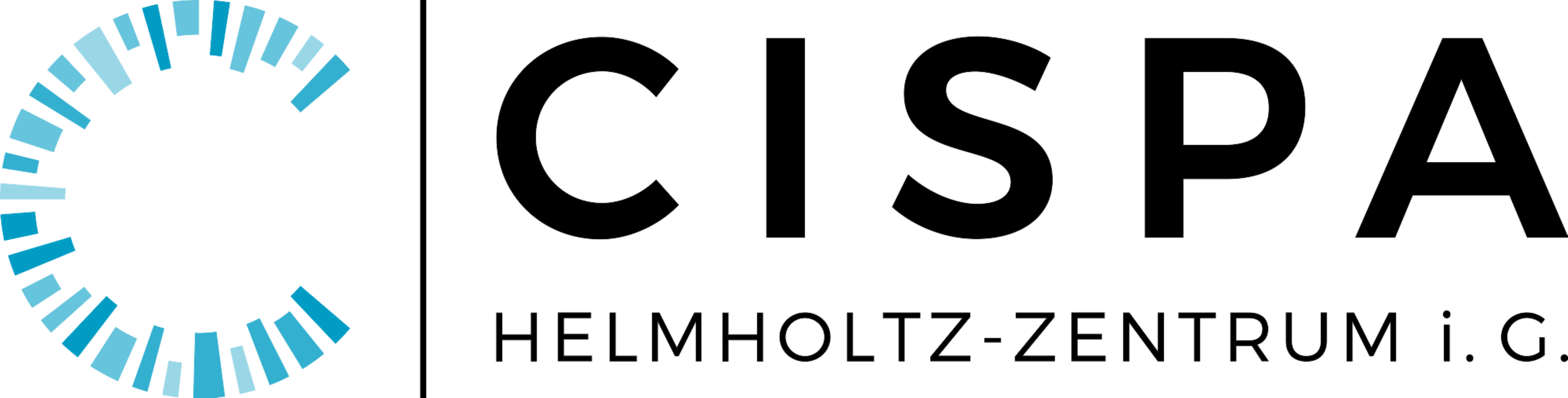}}
}
\usepackage{csquotes}
\usepackage{etoolbox}

\usepackage[nohead,includefoot,margin=2cm]{geometry}
\usepackage[compact,raggedright,small]{titlesec}
\usepackage[affil-it]{authblk}
\usepackage[runin]{abstract}
\usepackage{titling}
\usepackage{setspace}

\usepackage{multirow}
\usepackage{subcaption} %
\usepackage{caption}
\usepackage{tikz}
\usetikzlibrary{arrows}
\usetikzlibrary{positioning}

\usepackage{xspace}
\usepackage{hyperref}  %
\usepackage[nameinlink,noabbrev]{cleveref}  %
\usepackage[inline]{enumitem}
\usepackage{microtype} %
\usepackage{alltt}
\usepackage{float}
\usepackage{stfloats}

\usepackage[noend]{algpseudocode}
\usepackage{listings}
\usepackage{relsize} %

\definecolor{codegreen}{rgb}{0,0.6,0}
\definecolor{codegray}{rgb}{0.5,0.5,0.5}
\definecolor{codepurple}{rgb}{0.58,0,0.82}
\definecolor{backcolour}{rgb}{0.95,0.95,0.92}

\lstdefinestyle{mystyle}{
    commentstyle=\color{codegreen},
    keywordstyle=\color{magenta},
    numberstyle=\tiny\color{codegray},
    stringstyle=\color{codepurple},
    basicstyle=\footnotesize,
    breakatwhitespace=false,
    breaklines=true,
    captionpos=b,
    keepspaces=true,
    numbers=left,
    numbersep=5pt,
    showspaces=false,
    showstringspaces=false,
    showtabs=false,
    tabsize=2
}

\algblockdefx[switch]{Switch}{EndSwitch}%
[1]{\textbf{switch} #1}%
{}%
\algloopdefx{Case}[1]{\textbf{case} #1:}%
\algdef{SE}[DOWHILE]{Do}{doWhile}{\algorithmicdo}[1]{\algorithmicwhile\ #1}%

\def\|#1|{\textit{#1}}
\def\<#1>{\texttt{#1}}

\newcommand{\klee}{\textsc{Klee}\xspace}
\newcommand{\radamsa}{\textsc{Radamsa}\xspace}
\newcommand{\basilisk}{\textsc{Basilisk}\xspace}

\newcommand{\llvm}{\textsc{LLVM}\xspace}
\newcommand{\llvmir}{\textsc{LLVM~IR}\xspace}

\usepackage{framed}
\newenvironment{result}{\begin{framed}\centering\it}{\end{framed}}

\title{Carving Parameterized Unit Tests}
\date{\small (Dated \today)}
\author{Alexander Kampmann}
\author{Andreas Zeller}
\affil{\{kampmann, zeller\}@cispa.saarland \\
CISPA / Saarland University, Saarland Informatics Campus, Saarbr\"ucken, Germany}

\newcommand\BackgroundPic{
    \put(0,0){
    \parbox[b][\paperheight]{\paperwidth}{%
    \vfill
    \centering
    \includegraphics[width=\paperwidth,height=\paperheight]{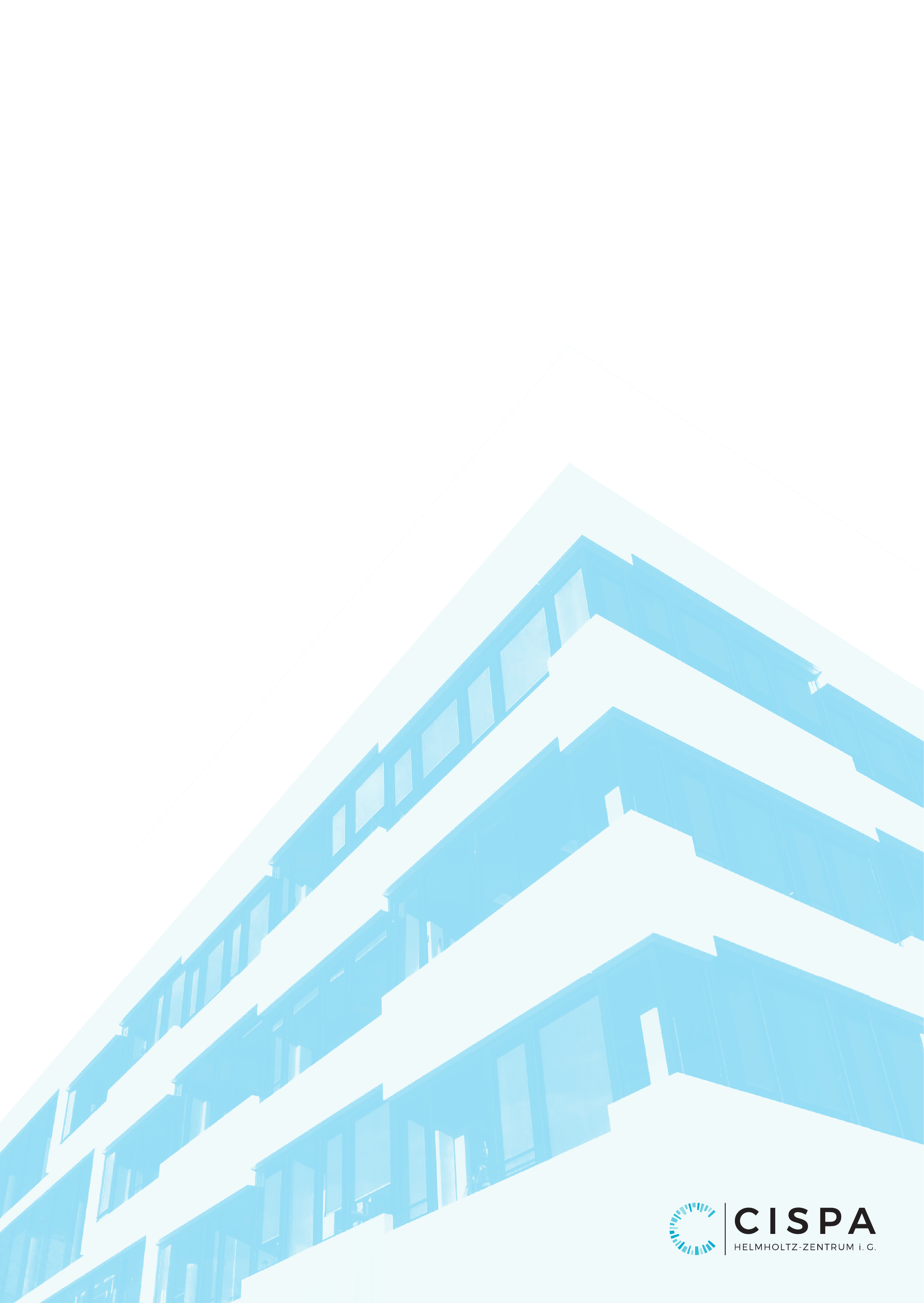}
    \vfill
}}}

\usepackage{utopia}
\usepackage{times}

\begin{document}
\AddToShipoutPicture*{\BackgroundPic}

\makeatletter
\renewcommand{\Authfont}{\normalsize\sffamily\bfseries}
\renewcommand{\Affilfont}{\normalsize\sffamily\mdseries}
\begin{titlepage}
\newcommand{\HRule}{\rule{\linewidth}{0.1mm}}
\centering
  \textsc{\LARGE {\fontfamily{Montserrat-TOsF}\selectfont CISPA Helmholtz-Zentrum i.G.}}\\[1.5cm]

  \vspace{2.4 cm}
  \HRule \\[0.2cm]
  {\huge\sffamily\bfseries \@title\par}
  \vspace{0.2cm}
  \HRule \\[1.5cm]

  {\sffamily \@author\par}
\vfill

\end{titlepage}

\makeatother
\setlength{\affilsep}{0.1em}
\addto{\Affilfont}{\small}
\renewcommand{\Authfont}{\normalsize}
\renewcommand{\Affilfont}{\normalsize}

\pretitle{\begin{center}\large\bfseries}
\posttitle{\end{center}}
\maketitle
\thispagestyle{CISPA}

\begin{abstract}

We present a method to automatically extract (``carve'') parameterized unit tests from system executions. The unit tests execute the same functions as the system tests they are carved from, but can do so much faster as they call functions directly; furthermore, being parameterized, they can execute the functions with a large variety of randomly selected input values.  If a unit-level test fails, we lift it to the system level to ensure the failure can be reproduced there.  Our method thus allows to focus testing efforts on selected modules while still avoiding false alarms: In our experiments, running parameterized unit tests for individual functions was, on average, 30~times faster than running the system tests they were carved from.
\end{abstract}

\section{Introduction}
\label{sec:introduction}

Tools and methods for software test generation can be distinguished by the \emph{level} at which they feed generated data into a program.  
At the \emph{unit level,} test generation operates by invoking individual functions, allowing for effectively narrowing down the scope of analysis and execution, while at the same time making internals directly available for testing.  The downside, however, is that synthesized function calls may \emph{violate implicit preconditions}: If a test generator finds that \<sqrt(-1)> crashes, this does not help developers who never intended \<sqrt()> to work with negative numbers anyway.
When generating tests at the \emph{system level,} this problem of false failures does not occur, as a system is expected to reject all invalid inputs; and any failure caused by third-party system input needs to be fixed.  On the other hand, system-level testing must read, decompose nad process inputs, before the function of interest is finally reached. This leads to \emph{overhead}, as compared to a unit-level test.  Furthermore, effective system test generation is often hampered by scale: symbolic analysis, for instance, hardly scales to system sizes.

In this paper, we present a method that joins the benefits of both system-level and unit-level test generation, while at the same time avoiding their disadvantages.  Our key idea is based on the concept of \emph{carving unit tests}~\cite{elbaum2009carving}, observing system executions to extract unit tests that replay the previously observed function executions in context.  However, we extend the concept by extracting \emph{parameterized unit tests}, allowing to replay not only the original function invocations, but also to synthesize several more.  To this end, we identify those function arguments that are directly derived from system input.  These arguments then become unit test parameters, allowing for extensive fuzzing with random values.  We can thus random test individual functions with hundreds of values, with all invocations in the context of the original run, but without requiring the overhead of starting the program anew for each system test run.  Furthermore, we can generate tests for any subset of unit tests as we like, spending testing time on error-prone or recently changed functions.

\begin{figure}
	\includegraphics*[width=\linewidth]{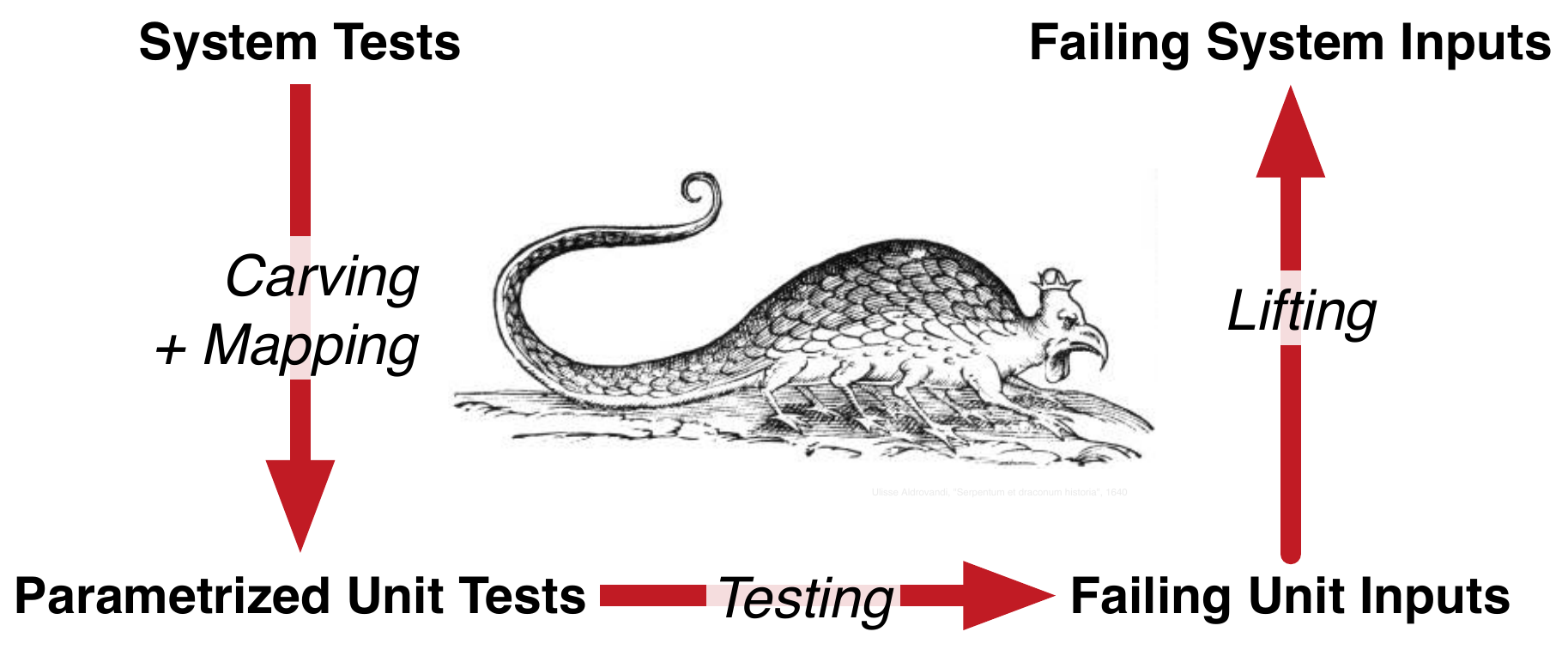}
	\vspace{-1.25\baselineskip}
	\caption[Overview of our approach.]{Overview of our approach.   Starting from a set of (given or generated) system tests, our \basilisk prototype extracts (``carve'') parameterized unit tests, each representing a function with symbolic parameters in the context observed during system testing.  The unit tests are then exhaustively tested.  Arguments found to cause failures at the unit level are then lifted back to the system level, using a previously established mapping between function arguments and identical values occurring in the system input; the new system test is re-run to ensure it reproduces the failure.  The carved parameterized unit tests can be obtained from arbitrary system test generators, and be explored with arbitrary unit test generators.}
	\label{fig:overview}
	\vspace{-0.5\baselineskip}
\end{figure}

As an example of a carved parameterized unit test, consider the function \<bc\_add()> from the \<bc> calculator program.  \<bc\_add()> accepts two numbers, \<n1> and \<n2>, encoded in \<bc>'s internal representation for numbers with arbitrary floating-point precision, and writes the sum of those two numbers to the number pointed to by \<result>. \<scale\_min> gives the minimal number of floating-point positions to be used by \<result>.
\begin{lstlisting}
void bc_add (bc_num n1, bc_num n2, bc_num *result,
int scale_min);
\end{lstlisting}
From an execution of \<bc> with a concrete input (say, \<"1 + 2\char34{}>), our \basilisk implementation observes the call \<bc\_add(1, 2, \&result, 0)>.  It identifies \<1> and \<2> as coming from system input, and thus makes them parameters of the carved parameterized unit test \<test\_bc\_add()>:
\begin{lstlisting}
void test_bc_add(int p1, int p2) {
// set up the context
bc_num n1;
bc_int2num(&n1, p1);
bc_num n2;
bc_int2num(&n2, p2);

// call the function under test
bc_num result;
bc_add(n1, n2, &result, 0);
}
\end{lstlisting}
We can now call \<test\_bc\_add()> with random values for~\<p1>~and~\<p2>, thus testing it quickly without having to start the \<bc> program again and again:

\begin{lstlisting}
test_bc_add(337747944, 352295539);
test_bc_add(535612873, 790525737);
// ... and more
\end{lstlisting}

Feeding random values into a function call brings the risk of violating implicit preconditions, even in the context of a concrete run.  In the case of a function failing, we thus \emph{lift} the failing unit test back to the system level; we can do so because we know where the original function argument came from in the system input.  Only if the failure can also be repeated at the system level do we report the failure.  Lifting is also useful if one is interested in \emph{coverage:} If a unit test achieves new coverage, we lift it to the system level, and verify whether the new coverage also applies there.

In our example, let us assume that \<test\_bc\_add(10, 20)> fails.  From the original run, we know that the arguments \<p1> and \<p2> correspond to the values  \<1> and \<2> in the system input.  In the input, we would thus replace the values \<1> and \<2> by the failure-inducing values of \<p1> and \<p2>, resulting in the input \<10 + 20>.  Only if \<bc> fails on this input would we report the failure. The lifting process thus gives us (1)~system inputs that fail (or obtain coverage), (2)~the same zero false positive rate as system tests, yet (3)~the speed and convenience of unit-level testing.

The overall interplay of carving, testing, and lifting is sketched in \Cref{fig:overview}.  As our approach is independent from a specific system-level or unit-level test generator, it allows to combine arbitrary system-level test generators (random, mutation-based, grammar-based, \dots) with arbitrary unit-level generators (random, symbolic, concolic, \dots).  In this paper, we explore one such combination, paving the way for many more that bring the best of both worlds.

The remainder of this paper is organized as follows.  After discussing the background (\Cref{sec:background}), this paper details its individual contributions:
\begin{enumerate}
	\item We present our approach for carving unit tests out of C~programs (\Cref{sec:carving}).
	\item We show how to identify \emph{parameters} from system input, thus carving \emph{parameterized unit tests} (\Cref{sec:parameterizing}).
	\item We demonstrate how to fuzz C programs with random values in a carved context (\Cref{sec:fuzzing}).
	\item We show how to \emph{lift} failed unit tests to the system level for validation (\Cref{sec:lifting}).
\end{enumerate}
In \Cref{sec:evaluation}, we evaluate \basilisk on a series of C~programs.  We find that carving parameterized unit tests can yield more coverage in less time when compared against system-level testing; these savings are multiplied when focusing the testing effort on a subset of the program.  \Cref{sec:conclusion} closes with conclusion and future work.

\section{Background}
\label{sec:background}

\subsection{Carving Unit Tests}

Carving unit tests was introduced by Elbaum et al.~\cite{elbaum2009carving} as a means to speed up repeated system tests, for instance in the context of regression testing.  Elbaum's work used the Java infrastructure, serializing and deserializing objects to enable faithful reproduction of units in context.

In contrast to this, our implementation carves unit tests from C programs. Instead of serialization, we use a heap traversal to record heap structures, and generate code to recreate a similar heap. To the best of our knowledge, \basilisk~is the first tool to implement any kind of unit test carving for~C.

Also we extended the carving mechanism to carve \emph{parameterized unit tests}~\cite{tillmann2005parameterized}, where function parameters whose values can be mapped to system input are left as parametric and thus open for unit test generators to explore.

\subsection{Extracting C Memory Snapshots}

Interpreting and recovering C data structures at runtime is notoriously difficult, since every programmer can implement not only her own data structures, but also her individual memory management. The work of Zimmermann and Zeller~\cite{Zimmermann2001graphs} on extracting and visualizing C runtime data structures (``memory graphs'') as well as their later application in  debugging~\cite{Zeller2002causeEffect, Cleve2005causes, Polishchuk2007inference} is related to ours in that all these works attempt to obtain a reliable and reproducible snapshot of C data structures.  Yet, all these works use these data structures for the purposes of debugging and program understanding rather than carving.

\subsection{Generating System Tests}

The idea of \emph{generating software tests} is an old one: To test a program~$S$, a producer~$P$ that will generate inputs for~$S$ with the intent to cause it to fail.  To find bugs, a producer need not be very sophisticated; as shown in the famous ``fuzzing'' paper of~1989, simple random strings can quickly crash programs~\cite{miller1990}.  

To get deeper than scanning and parsing routines, though, one requires syntactically correct inputs.  To this end, one can use formal specifications of the input language to generate inputs---for instance, leveraging \emph{context-free grammars} as producers~\cite{purdom1972}.  The \textsc{Langfuzz} test generator~\cite{Holler2012langfuzz} uses its grammar for \emph{parsing} existing inputs as well and can thus combine existing with newly generated input fragments.

Today's most popular test generators take \emph{input samples} which they mutate in various ways to generate further inputs.   \emph{American Fuzzy Lop,} or AFLFuzz, combines mutation with search-based testing and thus systematically maximizes code coverage~\cite{zalewski2018aflfuzz}.  More sophisticated fuzzers rely on \emph{symbolic analysis} to automatically determine inputs that maximize coverage of control or data paths~\cite{Godefroid2008whitebox}. The \klee~tool~\cite{Cadar2008klee} is a popular symbolic tester for C~programs. 

In our experiments, we use \radamsa~\cite{helin2018radamsa} which applies a number of mutation patterns to systematically widen the exploration space from a single input.  Instead of \radamsa, any other system-level test generator could also do the job.

\subsection{Generating Unit Tests}

The second important class of testing techniques works at the \emph{unit level,} synthesizing calls of individual functions.  These techniques separate in two branches: \emph{random} and \emph{symbolic}.  

\emph{Random} tools operate by generating random function calls, which are then executed.  A typical representative of this class is the popular \textsc{Randoop}~\cite{pacheco2007randoop} tool.  Random calls can be systematically refined towards a given goal: EvoSuite~\cite{Fraser2011evolution} uses a search-based approach to evolve generated call sequences towards maximizing code coverage.

\emph{Symbolic} techniques symbolically solve \emph{path conditions} to generate inputs that reach as much code as possible. PEX~\cite{tillmann2008pex} fulfills a similar role for .NET programs, working on \emph{parametrized unit tests} in which individual function parameters are treated symbolically.

Compared to the system level, test generation at the unit level is very efficient, as a function call takes less time than a system invocation or interaction; furthermore, exhaustive and symbolic techniques are easier to deploy due to the smaller scale.  The downside is that generated function calls may lack realistic context, which makes exploration harder; and function failures may be false alarms because of violated implicit preconditions.  Our parameterized unit tests supply a realistic context for unit-level testing; also, validating all unit-level failures at the system level means that we can recover from false alarms and any remaining failures are true failures.

\subsection{Generating Parameterized Unit Tests}

A number of related works has focused on obtaining parameterized unit tests by starting from existing or generated unit tests.  Retrofitting of unit tests~\cite{Thummalapenta2011retrofitting} is an
approach where existing unit tests are converted to parameterized unit
tests, by identifying inputs and converting them to parameters.  The technique of Fraser and Zeller~\cite{Fraser2011gpu} starts from concrete inputs
and results, using test generation and mutation to systematically
generalize the pre- and postconditions of existing unit tests.  The recently presented \emph{AutoPUT} tool~\cite{Tsukamoto2018autoPUT} generalizes over a set of related unit tests to extract common procedures and unique parameters to obtain parametrized unit tests.  In contrast to all these works, our technique carves parameterized unit tests directly out of a given run, identifying those values as parameters that are present in system input.

\section{Carving Unit Tests}
\label{sec:carving}

\subsection{How it Works}

The concept of \emph{carving} was introduced by Elbaum et al. as a general way to generate unit-tests out of system tests~\cite{elbaum2009carving}. Conceptually, a system test can be represented as the execution path through the program under test that is taken when the program processes the test inputs. In carving, we select one function invocation, that is, a subpath which starts with the invocation of a function $f$, and ends when $f$ returns. We call $f$~the~\emph{function (or unit) under test}. We also record the values of all global variables, as well as all parameters for this invocation of~$f$. We call the set~$C$ of recorded variable and parameter values the context. For the context~$C$, it is necessary to record heap structures. Variables or function parameters may be pointers into the heap; the heap itself may again contain pointers, forming a (potentially large) structure of heap objects. 

Just like regular unit tests, carved unit tests consists of three parts:
\begin{itemize}
	\item \textbf{Setup.} The setup code populates all variables with the values we recorded in~$C$. It reconstructs all heap structures recorded in $C$. 
	\item \textbf{Test.} The test part invokes the function~$f$, using the values that were constructed in the previous step as parameters. 
	\item \textbf{Tear down.} The tear down step releases all resources that were acquired by the setup. 
\end{itemize}
If the function~$f$ uses global variables and parameters only, the test is deterministic.

\subsection{Example}

Let us again take a look at the \<bc> calculator.  When \<bc> parses the input \<"1 + 2\char34{}>, each number is then stored in \<bc>'s internal representation, as a \<bc\_num> structure. Then, \<bc\_add()> is invoked with two \<bc\_num>'s, one representing \<1>, and the other representing \<2>. This leads to the carved (non-parameterized) unit test as follows:

\begin{lstlisting}
void test_bc_add() {
// set up the context
bc_num n1;
bc_int2num(&n1, 1);
bc_num n2;
bc_int2num(&n2, 2);

// call the function under test
bc_num result;
bc_add(n1, n2, &result, 0);
}
\end{lstlisting}

\subsection{Implementation}
\label{sec:carving-implementation}

In theory, building a carving tool is easy. Just take all variables and their values and write them out. The practice of carving, and especially the practice of carving for C~programs is a challenge.

Our \basilisk prototype implements carving based on the \emph{low-level virtual machine} (\llvm)\cite{lattner2004LLVM}. \llvm~provides a compiler, \emph{clang,} which compiles C code to an intermediate representation (\llvmir). Clang also compiles \llvmir~to machine-code. 

The \llvmir intermediate representation is designed to be used by compiler optimizations, that is, static analysis of the code. It removes all the syntactic sugar that was added to C for the sake of human developers, so it is hard to read for humans, but better suited for automatic analysis. At the same time, it pertains type information, which makes an analysis simpler than analyzing machine code directly. 

\basilisk works in two phases. It statically instruments the \llvmir~code, inserting probes which report all method invocations including the parameters, as well as all writes to global variables. During execution, those probes write the observed values at those points to a trace file.

For primitive types, like ints or floats, observed values can be written directly. However, there are more challenging situations for other data types, as listed below. 

\subsubsection{Pointers}

In \llvmir, as in C, a pointer is no more than a memory address. There is no information about the length of the memory segment that a pointer points to. But then, what should we dump out? 

Dumping only the byte that the pointer points to would mean we do not see the remainder of the memory area. Keep in mind that pointers are often used as arrays, e.g. a \<i8*> pointer, \llvmir~for a \<char *> pointer, is often used to point to a string, an array of characters in memory. In this case, dumping just one byte would give only the first character of the string. 

At the same time, we can not just dump arbitrary amounts of data. We would dump data that does not belong to this variable, and in some cases, accessing memory past the end of an allocated memory segment could even trigger a segmentation fault, and thereby crash the program under test. 

To solve this problem, we maintain a map that records, for each pointer, the length of the memory segment it points to. Keep in mind that pointers can be calculated from other pointers. If a pointer \<a> points to a memory segment of length \<10>, \<b = a + 5> gives a pointer \<b> which points to a memory segment of length 5, the second half of the segment pointed to by \<a>. The map needs to be able to identify \<b> as pointing into the memory segment at \<a>, and calculating the remaining length \footnote{Just in case you thought you could use an associative map for this lookup}. This allows us to find out how much memory should be dumped for a pointer. 

In the following, we explain how to populate the map. There are two kinds of pointers:

\begin{description}
	\item[Stack pointers] point to local variables or function arguments. The memory areas for those objects are part of the stack, they are allocated (and deallocated) by the runtime when a function is called (or returns). 
	As the compiler needs to be able to generate code to allocate a new stack frame, the compiler is capable of calculating the size of those memory areas. Our instrumentation gets the size at instrumentation time, the concrete pointer is captured at runtime, and written to the map. 
	\item[Heap pointers] point to memory on the program heap. In C (and in \llvmir, if the C standard library is used), memory is allocated with the \<malloc()> function. \<malloc()> receives the required length for a memory segment, and returns a memory segment of this length. Tracking all calls to \<malloc()>, we can update the map on each call to \<malloc>.
\end{description} %

\subsubsection{Strings}

For strings, one might assume that we can rely on the fact that they are zero-terminated, that is, the last character will be zero. Unfortunately we can not. Zero-termination is a convention, not a rule. Programmers may also decide to accompany their \<char*> variables with an integer variable, holding the length of the string. Also, we encountered several cases where a bitset was stored in a \<char *>. In a bitset, there may be relevant data behind a zero byte. So assuming zero-termination in those cases means that we would loose data. 

We thereby decided to handle \<char *> in the same way as any other pointer type. 

We saw that strings have uninitialized data at the end quite often, so if a \<char *> was dumped, we would, in constructing the context, also try to construct the context with the assumption that the string was zero-terminated. In many cases, this removed uninitialized segments at the end of the string. 

\subsubsection{Structs and Unions}

Struct types and union types are derived types. For a struct, that means that the struct consists of several values of different types, written to memory one after another. 
A union describes several options on how to interpret the same memory segment, e.g. 4-bytes of data may either be a single int32 value, or a four-letter string. 

We dump structs by handling each field recursively. Unions are compiled to structs and bitcasts (reinterpreting bytes in memory as some other type) in \llvm, so we don't need to think about how to handle them, \llvm already did.

\subsubsection{Extensive lengths}

For some subjects, we encountered large heap structures. Those might be large arrays, either as a \llvm array type or with a pointer that points to a lot of memory, or large connected structures, e.g. hash tables with lots of entries and long bucket lists. This is a problem, because the runtime overhead of dumping such large structures is extreme and a unit test which has to reconstruct such a structure will be large and slow. As our evaluation will show, slow unit tests are a problem for our approach. 

We solved the problem by introducing a size limit for the heap structures we dump. If a structures is too large, we simply abort dumping it. This means that the carver has to deal with incomplete structures, and some unit test may not build a complete context. However, as the lifting step filters false alarms, this does not pose a big problem. 

\subsubsection{External Resources}
If the program under test had a file open when $f$ was invoked, reconstructing variable values and heap structures will not be sufficient. Calls to \<read()> or \<write()> on the file will fail.  Similar situations may occur with other resources, such as locked mutexes or open network connections.  

A perfect solution would have to deeply interact with the operating system (and the system environment) to perfectly preserve and reconstruct states, which is out of our scope. We thereby just ignored the issue. It may lead to false alarms in the unit tests, but our lifting step will filter those. 

\subsubsection{Writing to Global Variables}

We also want to dump the values of global variables. However, when a function is called we can not know which global variables it uses. Thereby we need to dump new values for global variables whenever a global variable is written to. 

This is rather easy. \llvmir uses the \<store> instruction to write to memory. If \<store> writes to a global variable, we dump the new value for this variable. 

Unfortunately, this is not enough. If the global variable is a struct or array, individual members may be written. In this case, \llvmir~uses the \<getelementptr> instruction to calculate the address of this member first. For an array of structs, or struct members in a struct, the result of \<getelementptr> may be used in an address calculation again. If the global variable is a pointer to a pointer, the \<load> instruction may be used to retrieve the underlying pointer. 

We solved this as follows: When our instrumentation encounters a \<store> instruction, it traces back the pointer operand, until it hits a value which is not a \<getelementptr> and not a \<load>. If this value is a global variable~$v$, the \<store> is considered to be a write to~$v$ and the value is dumped. 

\subsubsection{Program-Specific Extensions}
Even with all the above heuristics, programmers may still choose internal data representations that make it hard for \basilisk to recognize data.  As an example, \<bc> uses char arrays (strings) to represent numbers as a string of digits; However, they do not use ASCII-encoded digits, which our heuristics for strings could handle, but the integer values of the characters directly. 

To address this issue, we implemented special handling for the arbitrary-precision numeric types in \<bc>.  This special handling is just 76 lines of kotlin code. More careful engineering of our prototype could make this even easier. This special handling allowed us to use the \<bc\_int2num()> function from the subject to set up the environment. 

Other than that, \basilisk was able to precisely identify, extract and reproduce almost all data structures for all of our subjects.  Still, depending on how creative programmers are as it comes to their own memory management, similar extensions may be required.

\section{Parameterizing Carved Unit Tests}
\label{sec:parameterizing}

\subsection{How it Works}

We want to use our extracted unit tests for test generation.  Therefore, they need to be \emph{parameterized}---that is, there need to be parameters whose values can be set by the fuzzer.

In principle, the fuzzer may change any value in~$C$. However, this bears the problem of generating \emph{false positives.}  A failure of a function under test is only relevant if it can be triggered with system-level inputs.  A failure due to invalid unit-level inputs is irrelevant if those inputs will never be provided to the unit in a system-level invocation.  We therefore restrict ourselves to \emph{values that are derived directly from system-level input.}

System-level input consist of command-line arguments, input files, and other inputs. For ease of presentation, we will assume that all inputs are just one string~$S$.  In order to identify parameters that would be directly derived from~$S$, we need the \emph{data flow} from the system-level input~$S$ to the unit-level input~$C$, such that we can identify the \emph{origin} of each and every variable in~$C$.  To establish such a data flow, we could use \emph{dynamic tainting}, tracking all input characters throughout the execution as well as their derived values, and eventually checking which of these reach the variables in~$C$.  However, we are interested only in \emph{direct} flows that can be easily inverted---that is after a change in $C$, we want to be able to easily generate a system input that represents the change as well.  Furthermore, dynamic tainting can be very slow, in particular considering that one may be interested only in a small set of functions and their arguments.

To match variables and their origins in the system input, we thus use a simple, yet efficient approximation.  We traverse the variable values and heap structures in $C$. For each $v \in C$, we check whether~$v$ occurs in~$S$:
\begin{itemize}
	\item
	If $v$ is a string, we check whether it is a substring of $S$.
	\item
	If~$v$~is~numeric, an integer or a floating-point number, we check whether the decimal representation is a substring of~$S$.
\end{itemize}
If we find a match, we mark~$v$~as~a~parameter. Instead of using the recorded~value~$v$, we now allow the fuzzer to insert a new~value~$v'$ into the unit test, as a replacement for~$v$.

\subsection{Example}

In our running example, \basilisk identifies the value~\<1> of the first method parameter as being related to characters 0~to~1 of the system-level input \<"1 + 2\char34{}>. The second argument~\<2> is related to characters 4~to~5 of the same input. Thereby it turns those two values into parameters. The fuzzer may now choose new values for those parameters. 

\subsection{Implementation}
\label{sec:reachable}

As described before, we check whether a method parameter or the value of a global variable is a substring of one of the system-level inputs. We handle $S$ as a set of values per input source, and we do the string comparisons for each input source individually. 

For \<char *> variables, we interpret them as strings and check them for substrings. This may not always be the correct approach, because as described above, \<char *> variables are not always strings; yet, this brought sufficient results.

For integer variables and variables with a floating-point type, we used the usual decimal encoding to convert them to strings and again applied a substring comparison. This may limit the applicability to other subjects. If a subject accepts numeric input in, e.g. hexadecimal encoding, our prototype would not detect that and thereby miss opportunities to symbolize parameters. 

Global variables frequently contain some system input. However, if this global variable is not used in the function under test, the unit test does not need to be parameterized. The values from the fuzzer would never be used. Thereby, we only consider reachable globals for parameterization. 

We consider a global variable as \emph{reachable} if there is a load from or a store to this variable in a reachable function. We consider a function as reachable if it is the function under test, called in a reachable function, if a function pointer to this function is created somewhere in a reachable function, or if a function pointer to the function exists in~$C$. 

We generate \llvmir~code which sets all global and local variables as recorded in~$C$, and calls the function~$f$.

\section{Fuzzing Function Calls}
\label{sec:fuzzing}

\subsection{How it Works}

Once we have a parameterized unit test, we can use a fuzzer to choose new values for those parameters. A fuzzer basically provides random values. 

Instead of simple random fuzzing, we could also use more sophisticated test generators; A tool like PEX~\cite{tillmann2008pex}, for instance (if it were available for~C), could apply symbolic constraint solving to systematically explore paths in the carved parameterized unit test.  Since the scope of the symbolic analysis would be constrained to only the carved parameterized test and the function under test, it would not suffer from the problems of scaling one would have when applying it at the system level.  Hence, our approach effectively enables a fusion of system-level and unit-level test generation.

\subsection{Example}

For the running example, the fuzzer generates invocations such as: 

\begin{lstlisting}
test_bc_add(337747944, 352295539);
test_bc_add(535612873, 790525737);
// ... and more
\end{lstlisting}

\subsection{Implementation}

We wrote a simple unit-level fuzzer to generate new values for all parameters. We execute the unit tests with those new parameters and report all cases where the unit test fails or covers previously uncovered code. 

For integer and double types, our fuzzer uses bitflips, random values and the values \<0>, \<INT\_MAX> and \<INT\_MIN>. For strings, our fuzzer uses bitflips, sequences of random bytes, sequences of random ASCII characters, all 0 strings and all \<0xFF> strings. Also, we implemented a mutator which takes the original string and repeats sequences thereof. 

Carving from a program run generates one (parameterized) unit test for each function that was called in the execution.  
\label{sec:selection}
The system maintains a list of all unit tests. When it is time to execute the fuzzer on a unit test, it orders the unit tests by the number of inputs that were already given to the function under test in other unit tests, and by the coverage all system tests known so far achieved on the function under test. The function with the lowest number of invocations, and among those with the same number of invocations the function with the lowest coverage so far will be parameterized and fuzzed first.

\section{Lifting Failure-Inducing Values}
\label{sec:lifting}

\subsection{How it Works}

As stated in the introduction, invoking functions with generated values runs the risk of \emph{false alarms,} with the values violating implicit preconditions for the usage of the functions.  Typical examples include numerical values that are out of range, strings with invalid contents, and more.  From the perspective of the function alone, one cannot distinguish whether the function failed because of a bug, or whether it failed because of an invalid input.  Only in the context of the whole system can one decide whether a failure is a true failure, because a system is not supposed to fail with an internal error in the presence of invalid input.  

A similar problem occurs with \emph{coverage.} If we are interested in maximizing coverage, and we can generate a unit-level test that covers a new program structure, we want to validate this at the system level---because, again, the coverage achieved locally may not be easily feasible at the system level.

To address these problems, we thus \emph{lift} input values from the function level back to the system level. Unit-level input values are selected for lifting, if they trigger a failure on unit-level, or reach new coverage on unit-level. 

 For each parameterized unit test, the fuzzer generates new values $v'$ for all parameters $v$. Now, each $v$ was a substring of the system-level input $S$. We recorded which interval in $S$ $v$ corresponds to.  Thereby, we can derive a new system-level input $S'$ by replacing $v$ with $v'$ in $S$. 

\basilisk then invokes the program under test with the input $S$ and observes the outcome of the execution. Ideally, the system test executes the same execution path as the unit test did. However, there are three possible outcomes

\begin{itemize}
	\item If the invocation of the program under test produces the same failure as the unit test, we have a \emph{true positive}, and generated a new, valid, and bug-revealing system test. We also provide the information which unit test and which unit-level input triggered the problem, which tells the developer where she should start debugging. 
	\item If the invocation of the program under test covers the same code as the unit test, we have a \emph{true positive}. We generated a new, valid system test which achieves additional coverage. 
	\item If the system test does \emph{not} fail and does \emph{not} achieve the new coverage, as predicted by the unit test, we have a false positive. False positives occur because in the system context, the unit-level values do not reach the function unchanged, or not at all.  Also, the failure may not occur because the context~$C$ is not completely reconstructed, or because the substring relation between $v$ and $S$ was circumstantial.  Whatever the reason, there is no need for the human software developer to look into the failure, as it cannot be reproduced at the system level.	
\end{itemize}

For all reported failures, the developer gets a system test as well as a unit test that both faithfully reproduce the failure (or achieve the new coverage).  In case of failure, the developer can use the system test to assess which external circumstances lead to the error, and also to demonstrate that the failure is real; and she can use the unit test to debug the program in context, without having to step through the entire system test.

\subsection{Example}

In our running example, the first parameter was related to characters 0 to 1 of \<"1 + 2\char34{}>, and the second parameter was related to characters 4 to 5. Assuming that the invocation 

\begin{lstlisting}
test_bc_add(337747944, 352295539);
\end{lstlisting}

reveals a bug or at least provides additional coverage, we generate the system-level input \<337747944 + 352295539>.

\section{Evaluation}
\label{sec:evaluation}

In our evaluation, we attempt to answer the following research questions:
\begin{enumerate} 
	\item How do the \emph{unit tests} generated by \basilisk compare against system tests from \radamsa?
	\item How do the \emph{system tests} generated by \basilisk compare against system tests from \radamsa?
	\item Which \emph{time savings} are possible if one wants to focus on a subset of functions?
\end{enumerate}

\subsection{Subjects}

We applied our prototype implementation on seven subjects. 

Four of the subjects are part of \emph{GNU coreutils}, a collection of standard command line tools which is used, e.g. on Linux. The \<cut> program reads text from a file and outputs substrings, as specified by the user. The \emph{paste} program, also part of \emph{GNU coreutils}, can be used to merge lines from different text files. The \<tac> command reads a file and outputs it in reversed order. \<b2sum> computes a message digest, some kind of checksum, from an input file. 

\<sed> is a stream editor that applies a list of user-specified commands on its input and outputs the resulting text.

The remaining two subjects are the \<bc> and \<dc> programs. Both of them are programming languages with arbitrary-precision floating-point arithmetic. \<bc> uses a C-like syntax, meaning that the input \<"1 + 2\char34{}> prints \<3>, as one would expect. \<bc> also allows for variables and functions. \<dc> uses reverse polish notation, where the operator follows its operands: \mbox{\<"1 2 +\char34{}>} yields the output \<3>. \<dc> has registers, which can be used as variables.  \<bc> and \<dc> share their arithmetic code.

The subjects are listed in \Cref{tab:subjects}. We also report the lines of code for each subject. Especially for the programs from \emph{GNU coreutils}, the source code repositories do contain more code. We counted only lines of code that are in functions that are reachable from the main method of the respective program. We used the reachability analysis we described in \Cref{sec:reachable}.

\begin{table}
	\caption{Evaluation subjects}
	\label{tab:subjects}
	\centering
	\sffamily
	\rowcolors{1}{Apricot}{white}
	\begin{tabularx}{\columnwidth}{lrrX}
		\textbf{Subject} & \textbf{LoC} & \textbf{Functions}  & \\
		\hline
		b2sum & 1228 & 115 & checksum calculation \\
		paste & 662 & 79 & text processor \\
		tac & 987 & 111 & text processor \\
		bc & 3456 & 151 & arbitrary-precision calculator \\
		dc & 1997 & 136 & arbitrary-precision calculator \\
		cut & 1346 & 127 & text processor \\
		sed & 2715 & 215 & text processor \\
	\end{tabularx}
\end{table}

\subsection{Running \basilisk and \radamsa}

Our \basilisk prototype starts with a set of \emph{seed tests.} First of all, it uses \radamsa to generate 10~additional system tests per seed test. Then, it carves unit tests from all system tests. Afterwards a unit test is selected as described in \Cref{sec:selection}, and the process of parameterizing, unit-level fuzzing and lifting is applied to this test. Once this is done, the next unit test will be selected and processed.  Generated system tests are executed immediately, and the coverage they generate may already be used in the next selection step. 

In our experiments, we used a time limit. Our prototype runs each system test directly after creating it, so the output is coverage information. \radamsa only generates test inputs. That means that comparing directly is not fair. While, for our tool, the time limit includes test executions, \radamsa needs additional time to execute the generated system tests. 

In order to mitigate this difference, we had \radamsa generate system tests in batches of 10 tests each and executed each batch before generating the next one, until the time limit was exceeded. This means that for \radamsa, as for \basilisk, system test execution time is included in the time limit.

\subsection{Unit Tests vs. System Tests}
\label{sec:comparing-time}

In our first experiment, we compared the unit tests generated by \basilisk with system tests produced by \radamsa.  Experiments were ran on an Intel i7-2600 processor at 3.40GHz with 16~GB RAM on Linux. We measured real time elapsed for the programs instrumented to obtain coverage information; this instrumentation is responsible for the relatively long executions.\footnote{We implemented our own coverage tool, as for our experiments, we would require the ability (a) to fully trace executions and (b) to maintain that trace (and the coverage) even in the presence of failures; the tool is thus optimized for reliability rather than performance.  An industrial strength implementation for coverage only would require only an overhead proportional to the code size, as it would replace instrumented with uninstrumented blocks once covered.}
The time limit was 15~minutes for both tools.

We used one hand-written seed test, which was identical for both tools. In designing the seed tests, we used input examples that we found in online tutorials or the bug tracker of the respective programs.  We ran each experiment 5~times with different seeds; the results are means over all five runs.

\Cref{tab:speedup} lists the mean execution time for the \basilisk unit tests as well as the \radamsa system tests.  We see that, as expected, a single execution of a carved parameterized unit test with arguments runs much faster than a system test, up to a factor of~180 and with a mean factor of~30.

\begin{table}
	\caption{Execution times for system tests vs. unit tests}
	\label{tab:speedup}
	\sffamily
	\rowcolors{1}{Goldenrod}{white}
	\centering
	\begin{tabularx}{\columnwidth}{lrrr}
\textbf{Subject} & \textbf{Unit Tests} & \textbf{System Tests} & \textbf{Speed up} \\
\hline
b2sum   &             22.77ms &             1172.94ms &    51.50$\times$ \\
paste   &             33.74ms &             1836.20ms &    54.42$\times$ \\
tac     &             19.29ms &             3520.12ms &   182.44$\times$ \\
bc      &             87.37ms &             1023.46ms &    11.71$\times$ \\
dc      &            122.95ms &             1708.41ms &    13.90$\times$ \\
cut     &             41.38ms &              264.85ms &     6.40$\times$ \\
sed     &              6.82ms &              729.00ms &   106.85$\times$ \\
\hline
total & 47.76ms  &  1465.00ms & 30.76$\times$  \\
\end{tabularx}

\end{table}

\begin{result}
	Carved unit tests execute much faster than system tests.
\end{result}

The finding that unit tests execute much faster confirms the experiments of Elbaum et al., who reported that their carved tests ``reduce average test suite execution time to a tenth of our best system selection technique''~\cite{elbaum2009carving}.  In contrast to Elbaum et al., though, we are not limited to invocations seen during system testing, but can (and do!) generate additional ones.

\subsection{Overall Coverage}
\label{sec:overall-coverage}

Let us now see whether the unit test speedup brings benefits during testing.  To this end, we identify those unit tests that result in an increase in branch coverage.  We lift those tests to become system tests.  \Cref{tab:lifted_tests} shows that although \basilisk runs hundreds of thousands of unit tests, only a small fraction of these results in new branch coverage and yields new system tests.  While some of the unit tests fail, none of these failures still occur after lifting the generated arguments back to system tests.

\begin{result}
	Lifting unit test failures to system tests is effective in preventing false alarms.
\end{result}

\begin{table}
	\caption{Lifted unit tests}
	\label{tab:lifted_tests}
	\sffamily
	\rowcolors{1}{GreenYellow}{white}
	\centering
	\begin{tabularx}{\columnwidth}{lrrr}

\textbf{Subject} & \textbf{\#Unit Tests} & \textbf{\#Lifted Tests} & \textbf{\% lifted} \\
\hline
b2sum   &    545110.4 &         329.8 &              0.06\% \\
paste   &    219379.6 &         273.0 &              0.12\% \\
tac     &    872961.2 &          79.2 &              0.01\% \\
bc      &      8181.4 &         159.1 &              1.94\% \\
dc      &    396664.8 &         125.6 &              0.03\% \\
cut     &    909140.4 &         383.0 &              0.04\% \\
sed     &     25095.8 &         167.7 &              0.67\% \\
\hline
total & 2976533.6 & 1517.4 & 0.05\% \\
\end{tabularx}

\end{table}

If we measure the branch coverage of the system tests thus lifted, we can directly compare the coverage of \basilisk and \radamsa.  (At this point, both tools have spent the same test budget of 15~minutes, which for \basilisk includes carving, parameterization, and lifting.)  \Cref{tab:coverage} contrasts the number of tests produced and branch coverage achieved.

It is worthwhile to note that \basilisk achieves its coverage through \emph{fewer} tests than \radamsa. Such a lower number of system tests is helpful, as it (re-)executes faster. This is because \basilisk only generates system tests where the originating unit tests have achieved \emph{new} coverage (and where the coverage gain is confirmed for the system test after lifting), while \radamsa uses no such feedback from the program.  

Comparing the first column in \Cref{tab:coverage} and the second column in \Cref{tab:lifted_tests}, it can be seen that the number of tests that were generated in \basilisk's fuzzing stage, the difference between the columns, is small. So \basilisk mostly relies on lifting. \<cut> is an exception here. For \<cut>, many unit tests were selected for lifting (thus counted in \cref{tab:lifted_tests}), but could not be lifted (the coverage gain was not confirmed), so the number of system tests is lower than the number of lifting attempts. 

\begin{table*}
	\caption{Branch coverage achieved by \basilisk and \radamsa}
	\label{tab:coverage}
	\sffamily
	\rowcolors{1}{Goldenrod}{white}
	\centering
	\begin{tabular}{lrrrr}
{} & \multicolumn{2}{l}{\textbf{\#System Tests}} & \multicolumn{2}{l}{\textbf{Coverage}} \\
\textbf{Subject} & \textbf{\basilisk} &   \textbf{\radamsa} &      \textbf{\basilisk} &        \textbf{\radamsa} \\
\hline
b2sum   &     358.0 &   629.0 &    37.93\% &  19.49\% \\
paste   &     280.0 &   346.3 &    33.33\% &  31.08\% \\
tac     &      89.6 &   212.6 &    34.66\% &  30.71\% \\
bc      &     169.0 &   577.2 &    26.47\% &  28.46\% \\
dc      &     135.4 &   434.6 &    18.39\% &  41.06\% \\
cut     &     339.2 &  3117.1 &    21.00\% &  20.50\% \\
sed     &     175.7 &  1058.3 &    21.19\% &  15.73\% \\
\end{tabular}

\end{table*}

We see that on five out of seven subjects, \basilisk reaches a higher coverage than \radamsa.  The two subjects where \basilisk does not reach a higher coverage are \<bc> and \<dc>; for \<cut>, the gain is small as well.  Looking at the rightmost column of \Cref{tab:speedup}, these subjects have the smallest speed ups of unit tests in comparison to system tests.  The performance gained by executing a unit rather than the entire program is offset by the analysis time of carving and parameterization.  On the other hand, the subjects with a high speedup of unit tests vs. system tests can very much profit from the carved parameterized unit tests.

\begin{result}
	If the unit tests are sufficiently faster than the system tests,
	parameterized carved unit tests yield a higher coverage.
\end{result}

To paint a complete picture, let us also take a look at the functions invoked.  \Cref{tab:reached} compares \basilisk and \radamsa in terms of covered functions.

\begin{table}
	\caption{Number of functions reached by \basilisk and \radamsa}
	\label{tab:reached}
	\sffamily
	\rowcolors{1}{Apricot}{white}
	\centering
	\begin{tabular}{lrr}

\textbf{Subject} &  \textbf{\basilisk} &  \textbf{\radamsa} \\
\hline
b2sum   &         42 &       21 \\
paste   &         21 &       22 \\
tac     &         23 &       24 \\
bc      &         69 &       71 \\
dc      &         49 &       73 \\
cut     &         50 &       49 \\
sed     &         81 &       69 \\

\end{tabular}

\end{table}

The biggest difference between \basilisk and \radamsa is in the \<dc> subject.  Investigation of the generated tests shows that \radamsa manages to generate new operators, e.g. \mbox{\<"1 2 + p\char34{}>} may be mutated to \<"1 2 * p\char34{}>. \basilisk only mutated the numbers themselves. Thereby \radamsa~discovered more functions, namely those for multiplication and other operators, while \basilisk found more paths in the already discovered functions.  Of course, whether one or another strategy is more effective depends on the (typically unknown) distribution of bugs in practice.

\subsection{Focusing on Single Functions}
\label{sec:focused-coverage}

While our previous experiment focused on running \emph{all} functions, in practice, we typically want to focus on \emph{specific} functions.  For instance, one may wish to focus testing on recently changed functions, functions that have a history of failures, functions that are critical, or other reasons that demand extensive testing.

System test generators like \radamsa give the tester no means to focus test generation on specific functions.  With parameterized unit tests, as carved by \basilisk, this is easy: We just execute those unit tests that test the function of interest.

If we want to test a \emph{single} function, \Cref{tab:speedup} already gives us an indication of the speedup we can expect.  By executing only those unit tests related to a single function, we can expect a significant speed-up; in our experiments, this is the factor~30 already mentioned.  While this speedup occurs only after carving and parameterization, a high number of tests will amortize the effort for these steps.

\begin{result}
	In our experiments, after carving a parameterized unit test for a function, one can test the function on average 30~times faster than with the original system test.
\end{result}

This factor~30 we found in our experiments has to be taken with a grain of salt, as it will very much depend on the speed differences of individual functions vs. system execution as a whole.  If the function to be tested encompasses most of the functionality of the program (e.g., the \<main()> function in C), carving a parameterized unit test will not yield significant time advantages.  On the other hand, if a program takes multiple seconds to start up or to process inputs before it can call a function (think of starting a Web browser or an office program), the time savings factor can easily reach several orders of magnitude.  Such savings accumulate as the carved parameterized unit tests can be reused again and again.

\subsection{Focusing on Sets of Functions}
\label{sec:focus-functions}

Let us now assume we have not one, but a \emph{set} of ``focus functions'' we are especially interested in---a set that may also be reached by chance through unfocused testing.  \Cref{tab:focusfunctions} shows the focus functions we randomly chose per subject.

\begin{table}
	\caption{Focus functions for each subject}
	\label{tab:focusfunctions}
	\rowcolors{1}{Apricot}{white}
	\sffamily
	\centering
	
	\begin{tabularx}{\columnwidth}{lX}
		\textbf{Subject} & \textbf{Functions} \\
		\hline
		b2sum & blake2b\_init(), blake2b\_init\_param(), blake2b\_update(), blake2b\_final() \\
		paste & xstrdup(), xmemdup(), quotearg\_n\_style\_colon() \\
		tac & quotearg\_style() \\
		bc & lookup(), bc\_sub(), bc\_multiply(), bc\_out\_num(), bc\_add() \\
		dc & bc\_out\_num(), dc\_add(), dc\_mul(), dc\_multiply(), dc\_binop(), bc\_add() \\
		cut & xstrdup(), hash\_initialize(), xstrndup(), quote(), quote\_n(), c\_tolower() \\
		sed & compile\_string(), normalize\_text(), compile\_regex() \\
	\end{tabularx}
\end{table}

Again, we ran \basilisk for 15 minutes per subject.\footnote{All reported values are means over 5 runs.}  In the first ``unfocused'' setting, we ran \basilisk as described above, executing all unit tests without further discrimination.  In the second ``focused'' setting, we had \basilisk carve and explore parameterized unit tests \emph{only} for the focus functions.  \Cref{tab:focustime} shows the time spent in the focus functions in both settings.

\begin{table}
	\caption{Time in milliseconds spent in focus functions}
	\label{tab:focustime}
	\sffamily
	\rowcolors{1}{Goldenrod}{white}
	\centering
	\begin{tabular}{lrr}

\textbf{Subject} & \textbf{Unfocused} & \textbf{Focused} \\
\hline
b2sum   &    128.76 &  661.00 \\
paste   &    131.26 &  785.50 \\
tac     &     11.75 &  640.16 \\
bc      &    112.41 &  647.34 \\
dc      &    151.33 &  771.94 \\
cut     &     96.28 &  820.48 \\
sed     &     98.26 &  842.20 \\

\end{tabular}

\end{table}

We see that having \basilisk focus on a small set of functions executes these focus functions much more often than the unfocused version.  The time \basilisk spends on these functions increases by a factor of at least six.  In other words, we can test a function six times as much than with an unfocused setting, or we can test six times as quickly.

\begin{result}
	With carved parameterized unit tests, \\
	one can focus on a subset of functions to be tested, \\
	yielding significant time savings.
\end{result}

The differences are even more dramatic when comparing the \emph{number of invocations.}  In \Cref{tab:focuscount}, we see the number of invocations for the focus functions per configuration, now also including \radamsa.  We see that focusing on a subset of functions can yield savings of up to a factor of 18,000 (\<cut>).  Again, this factor depends on the average running time of a unit; for \<dc>, our focus functions do not yield savings over system testing, as they take too long to carve and run.

\begin{table}
	\caption{Number of invocations of focus functions}
	\label{tab:focuscount}
	\sffamily
	\rowcolors{1}{GreenYellow}{white}
	\centering
	\begin{tabular}{lrrr}
\textbf{Subject} & \textbf{Unfocused} & \textbf{Focused} & \textbf{\radamsa} \\
\hline
b2sum   &    140468 & 2167102 &    1114 \\
paste   &     24723 &  248898 &    1039 \\
tac     &        60 &    1938 &     172 \\
bc      &       498 &   96041 &   11131 \\
dc      &     19551 &   19388 &  259167 \\
cut     &    270514 & 2973383 &     163 \\
sed     &       562 &   69327 &    1119 \\
\end{tabular}

\end{table}

\begin{result}
	With carved parameterized unit tests, \\
	focus functions can be executed much more often.
\end{result}

\subsection{Discussion Summary}

In the end, speed gains through carved parameterized unit tests will depend on multiple factors: Even if we have no focus set at all and include the effort for carving, we may still see gains in coverage over time (and conversely, less time for the same coverage), as discussed in \Cref{sec:overall-coverage}.  On the other hand, the smaller the focus set, and the quicker the functions execute, the higher the gains will be---up to the dramatic speedups shown in \Cref{tab:speedup}, discussed in \Cref{sec:focused-coverage}.

\subsection{Limitations and Threats to Validity}
\label{sec:threats}

Like any empirical study, our evaluation is subject to threats to
validity, many of which are induced by limitations of our
approach. The most important threats and limitations are listed below.

\begin{itemize}
	\item Threats to \textbf{external validity} concern our ability to generalize the results of our study.  We cannot claim that the results of our experimental evaluation are generalizable. 
	A huge concern is that we establish mappings from inputs to unit-level values via (sub-)string equality. This approach will, most likely, fail for programs that do not process text, e.g. image analysis software.  Another concern is that, if individual units need a long time to execute, the gains through carved parameterized unit tests will be small compared to system testing; if individual units are too large, the overhead of analysis, carving, and parameterization may not be offset through faster unit test execution.  We counter the threat in making our research infrastructure available, allowing for replication and extension.
	
	As we detail in \Cref{sec:carving-implementation}, carving is inherently limited in its ability to reconstruct a given context, especially when including external resources.  A carved parameterized unit test involving external resources runs the risk of not being able to execute a function at all, and such failures will not be reported as they will not be reproducible in the lifted system test.  This can be partially amended through deeper interaction with runtime and operating system; but such extension is out of scope for the present paper.
	
	\item Threats to \textbf{internal validity} concern our ability to draw
	conclusions about the connections between our independent and
	dependent variables.  \radamsa is a random-driven test generator, yielding different results each time.  Since in our evaluation, \basilisk starts with the tests coming from \radamsa, and as it choses random values when fuzzing individual functions, we have a second influence of randomness.  Furthermore, the choice of the seed test(s) has a huge influence on the achieved coverage. A good seed test leads to more initial coverage, and thereby more carving opportunities, which benefits \basilisk. On the other hand, a good seed test also gives \radamsa more opportunities to mutate system-level inputs; these effects may level each other out.  We counter both threats by running the experiments multiple times with different seeds, reporting mean times across all runs.
	
	\item Threats to \textbf{construct validity} concern the adequacy of
	our measures for capturing dependent variables.  To evaluate the quality of our tests, we use standard measures such as code coverage and execution time, which are well established in the literature.
\end{itemize}

\section{Conclusion and Future Work}
\label{sec:conclusion}

Carved parameterized unit tests bring together the best of system-level and unit-level testing.  Like unit tests, they can be quickly executed and focused on a small set of locations; from system tests, they obtain valid and realistic contexts in which test generation takes place; and when a unit fails, the failure can be lifted to the system level and validated there, either suppressing a false alarm or yielding a failing system test.  As our evaluation shows, the greatest potential of carved parameterized tests is in the speedup they provide: Focusing on a small set of functions of interest allows to speed up testing by orders of magnitude when compared to system-level test generation.
On a more conceptual level, carved parameterized unit tests create a bridge between system-level test generators and unit-level test generators, which can be arbitrarily combined; it thus paves the way towards new and exciting combinations of the best of two worlds.

Although our present approach came to be by exploring and refining several alternatives, it is by no means perfect or complete.  Our task list for the future includes the following extensions:

\begin{itemize}
	\item \textbf{Dynamic tainting.}  Our method for associating unit-level values with system inputs works well for all our subjects; however, we would like to have a method that also works when the input undergoes a number of computation steps.  To this end, we want to apply \emph{dynamic tainting} to follow individual characters of a system input through an execution, along with their derived values; this would effectively allow to associate any function argument with the input subset that influenced it.  The downside is that dynamic tainting induces a massive overhead (which may be better spent on test generation), and that the transformation steps from system input to function argument may not be easily reversible, preventing the final lifting step.
	
	\item \textbf{Advanced unit testing.}  Rather than simply feeding random values into functions, we would like to apply symbolic or search-based test generators at the unit level such that we can cover functions in a guided fashion.  We are currently experimenting with the \klee~tool~\cite{Cadar2008klee} for this very purpose; first results show that its generated inputs provide higher coverage than our randomly produced inputs, but this advantage is offset by the time it takes for analysis.  Another tool on our list is \emph{libfuzzer}~\cite{libfuzzer}; this tool provides a sequence of random bytes, which it evolves according to the coverage achieved.
	
	\item \textbf{Advanced system testing.}  Instead of \radamsa, we plan to experiment with alternate system-level test generation tools to obtain a wider range of function calls and arguments.  Grammar-based testing, as in the \textsc{Langfuzz} test generator~\cite{Holler2012langfuzz}, would allow to quickly cover input features and thus functions and values; furthermore, we could directly associate grammar elements such as strings and numbers with function arguments.
	
	\item \textbf{Better carving.}  Carving is a bag of hurt, plain and simple.  We are happy we made it far enough to get the experiments running, and we think we do have a nice and clean carving infrastructure at this point.  There is definitely room for improvement, and we may work on this at some point; but if, in the meantime, you plan to implement a carving technique for C, please contact us: We are happy to share our carving infrastructure such that you can save a year or so of coding and another year of debugging.
\end{itemize}

To allow easy reproduction and validation of our work, we have created a replication package for download.  The package includes all raw experimental data, as well as \basilisk itself:

\begin{center}
	\url{https://tinyurl.com/basilisk-icse19}
\end{center}

\bibliography{main.bib}

\begin{thebibliography}{10}

\bibitem{libfuzzer}
{libFuzzer}: a library for coverage-guided fuzz testing.
\newblock https://llvm.org/docs/LibFuzzer.html.

\bibitem{Cadar2008klee}
Cristian Cadar, Daniel Dunbar, and Dawson~R. Engler.
\newblock {KLEE}: Unassisted and automatic generation of high-coverage tests
  for complex systems programs.
\newblock {\em Proceedings of the 8th USENIX conference on Operating systems
  design and implementation}, pages 209--224, 2008.

\bibitem{Cleve2005causes}
Holger Cleve and Andreas Zeller.
\newblock Locating causes of program failures.
\newblock In {\em Proceedings of the 27th International Conference on Software
  Engineering}, ICSE '05, pages 342--351, New York, NY, USA, 2005. ACM.

\bibitem{elbaum2009carving}
Sebastian Elbaum, Hui~Nee Chin, Matthew~B Dwyer, and Matthew Jorde.
\newblock Carving and replaying differential unit test cases from system test
  cases.
\newblock {\em IEEE Transactions on Software Engineering}, 35(1):29--45, 2009.

\bibitem{Fraser2011evolution}
Gordon Fraser and Andrea Arcuri.
\newblock Evolutionary generation of whole test suites.
\newblock {\em 2011 11th International Conference on Quality Software}, pages
  31--40, 2011.

\bibitem{Fraser2011gpu}
Gordon Fraser and Andreas Zeller.
\newblock Generating parameterized unit tests.
\newblock In {\em Proceedings of the 2011 International Symposium on Software
  Testing and Analysis}, ISSTA '11, pages 364--374, New York, NY, USA, 2011.
  ACM.

\bibitem{Godefroid2008whitebox}
Patrice Godefroid, Michael~Y Levin, David~A Molnar, et~al.
\newblock Automated whitebox fuzz testing.
\newblock In {\em NDSS}, volume~8, pages 151--166, 2008.

\bibitem{helin2018radamsa}
Aki Helin.
\newblock Radamsa.
\newblock https://gitlab.com/akihe/radamsa.

\bibitem{Holler2012langfuzz}
Christian Holler, Kim Herzig, and Andreas Zeller.
\newblock Fuzzing with code fragments.
\newblock In {\em Proceedings of the 21st USENIX Conference on Security
  Symposium}, Security'12, pages 38--38, Berkeley, CA, USA, 2012. USENIX
  Association.

\bibitem{lattner2004LLVM}
Chris Lattner and Vikram Adve.
\newblock {LLVM: A Compilation Framework for Lifelong Program Analysis \&
  Transformation}.
\newblock In {\em {Proceedings of the 2004 International Symposium on Code
  Generation and Optimization (CGO'04)}}, Palo Alto, California, Mar 2004.

\bibitem{miller1990}
Barton~P. Miller, Louis Fredriksen, and Bryan So.
\newblock An empirical study of the reliability of {UNIX} utilities.
\newblock {\em Commun. ACM}, 33(12):32--44, December 1990.

\bibitem{pacheco2007randoop}
Carlos Pacheco and Michael~D Ernst.
\newblock {Randoop:} feedback-directed random testing for {Java}.
\newblock In {\em Companion to the 22Nd ACM SIGPLAN Conference on
  Object-oriented Programming Systems and Applications Companion}, volume~2 of
  {\em OOPSLA '07}, pages 815--816, New York, NY, USA, 2007. ACM.

\bibitem{Polishchuk2007inference}
Marina Polishchuk, Ben Liblit, and Chlo\"{e}~W. Schulze.
\newblock Dynamic heap type inference for program understanding and debugging.
\newblock In {\em Proceedings of the 34th Annual ACM SIGPLAN-SIGACT Symposium
  on Principles of Programming Languages}, POPL '07, pages 39--46, New York,
  NY, USA, 2007. ACM.

\bibitem{purdom1972}
Paul Purdom.
\newblock A sentence generator for testing parsers.
\newblock {\em BIT Numerical Mathematics}, 12(3):366--375, Sep 1972.

\bibitem{Thummalapenta2011retrofitting}
Suresh Thummalapenta, Madhuri~R. Marri, Tao Xie, Nikolai Tillmann, and Jonathan
  de~Halleux.
\newblock Retrofitting unit tests for parameterized unit testing.
\newblock In {\em Proceedings of the 14th International Conference on
  Fundamental Approaches to Software Engineering: Part of the Joint European
  Conferences on Theory and Practice of Software}, FASE'11/ETAPS'11, pages
  294--309, Berlin, Heidelberg, 2011. Springer-Verlag.

\bibitem{tillmann2008pex}
Nikolai Tillmann and Jonathan de~Halleux.
\newblock Pex--white box test generation for {.NET}.
\newblock In Bernhard Beckert and Reiner H{\"a}hnle, editors, {\em Tests and
  Proofs}, pages 134--153, Berlin, Heidelberg, 2008. Springer Berlin
  Heidelberg.

\bibitem{tillmann2005parameterized}
Nikolai Tillmann and Wolfram Schulte.
\newblock Parameterized unit tests.
\newblock In {\em Proceedings of the 10th European Software Engineering
  Conference Held Jointly with 13th ACM SIGSOFT International Symposium on
  Foundations of Software Engineering}, ESEC/FSE-13, pages 253--262, New York,
  NY, USA, 2005. ACM.

\bibitem{Tsukamoto2018autoPUT}
Keita Tsukamoto, Yuta Maezawa, and Shinichi Honiden.
\newblock {AutoPUT}: An automated technique for retrofitting closed unit tests
  into parameterized unit tests.
\newblock In {\em Proceedings of the 33rd Annual ACM Symposium on Applied
  Computing}, SAC '18, pages 1944--1951, New York, NY, USA, 2018. ACM.

\bibitem{zalewski2018aflfuzz}
Micha\l{} Zalewski.
\newblock american fuzzy lop.
\newblock http://lcamtuf.coredump.cx/afl/.

\bibitem{Zeller2002causeEffect}
Andreas Zeller.
\newblock Isolating cause-effect chains from computer programs.
\newblock In {\em Proceedings of the 10th ACM SIGSOFT Symposium on Foundations
  of Software Engineering}, SIGSOFT '02/FSE-10, pages 1--10, New York, NY, USA,
  2002. ACM.

\bibitem{Zimmermann2001graphs}
Thomas Zimmermann and Andreas Zeller.
\newblock Visualizing memory graphs.
\newblock In {\em Revised Lectures on Software Visualization, International
  Seminar}, pages 191--204, London, UK, UK, 2002. Springer-Verlag.

\end{thebibliography}
\bibliographystyle{plain}
\end{document}